# AN EXPLORATION OF FACTORS INFLUENCING THE ADOPTION OF ICT ENABLED ENTREPRENEURSHIP APPLICATIONS IN NAMIBIAN RURAL COMMUNITIES

*Elizabeth Ujarura KAMUTUEZU [1], Heike WINSCHIERS-THEOPHILUS [1], Anicia PETERS [2]*

[1]*Namibia University of Science and Technology, 13 Jackson Kaujeua Street, Windhoek, 9000, Namibia,*

*eukamutuezu@gmail.com & hwinschiers@nust.na*

[2] *University of Namibia, 340 Mandume Ndemufayo Ave, Pionierspark, Windhoek, 9000,*

*Email: anicia.peters@gmail.com*

**Abstract**: Digital services have the potential to improve rural entrepreneurs' access to wider markets and increase their competitiveness among other benefits. Moreover, during the ongoing Covid-19 pandemic in which movement and physical contacts have been limited, businesses relied much on digital services. However, many Namibian rural entrepreneurs have not been able to use digital services to maintain their livelihood. Therefore, this study investigated the factors affecting the adoption of ICT enabled services by rural entrepreneurs. The study applied a cross-sectional survey of 77 respondents comprising 14 rural entrepreneurs and 63 rural community members from four sites. It was found that the five main factors affecting the adoption of digital services by rural entrepreneurs are a lack of awareness of digital services, electricity, skills to navigate smart devices, high cost of both devices and mobile internet and cybercrime. We recommend a tailor-made training program for rural entrepreneurs which includes raising awareness of digital services and associated benefits, capacity building on digital skills and best practice for cybersecurity. In addition, we propose that the Namibian Government should enhance digital inclusion through a policy initiative to reduce the cost to make both data and smart devices affordable for the poor and rural communities.

# 1. INTRODUCTION

ICT became a pillar of socio-economic development worldwide and a necessity during the pandemic. The inequalities of access to digital services and technologies have been felt more than ever. Many African countries have been battling with political uncertainty, economic adversity, social inequalities, urban-rural migrations, and underdeveloped technical infrastructures [20]. While there is an extensive body of literature focusing on the adoption of ICT by Small Medium Enterprises (SMEs) in the global south and governments recognising the significance thereof, little attention was given to rural areas in developing countries [1,9]. Moreover, many studies were conducted by researchers from developed countries with narrow exposure and lack of contextual understanding, mostly using positivist quantitative approaches with limited explanations of factors influencing ICT implementations [37]. While acknowledging the opportunities and benefits of ICTs, it is important to investigate the adoption levels, the barriers to adoption and devise alternatives to reduce the barriers in order to unpack ICT enabled services' contributions that improve people's lives. This study, therefore, aims to fill a research gap through an empirical investigation focusing on Namibian rural entrepreneurs. The aim of this study is to examine the factors influencing the adoption of ICT enabled applications and services for rural entrepreneurship activities in Namibia and recommends initiatives to overcome barriers to ICT adoption by rural entrepreneurs.

The Namibian Government has prioritised the deployment of ICT infrastructures to rural areas to bridge the digital divide between its urban (48%) and rural (52%) population. This is encapsulated in the broadband policy targeting 80% population coverage by 2021[26]. Namibia is boasting a 120% mobile phones penetration, which is higher than the World (67%) and Africa (81%) respectively and an internet connection of 51% while Africa has an average of 34%. Social Media usage has increased to 31% in 2021 [13]. COVID-19 has significantly increased mobile phone and internet usage across Namibia in many spheres of life. Due to lockdowns and limitations on movement, many people were communicating online, with the tertiary education sector nearly entirely transferred to an online teaching mode [10]. With the agricultural sector being a dominant player in the economy, the usage of ICT enabled services will play a significant role in promoting entrepreneurship and economic growth in rural areas by diversification of the rural economy embedded into competitive agriculture and forestry industries, and improved quality of life [21]. Generally, the high mobile penetration holds a prospect for Namibia to adopt ICT in rural areas [29] and yet during the pandemic, Namibian farmers were unable to travel to pay the workers or use online/cellular banking facilities to send the money to workers to buy the farming necessities. Thus, it needs to be determined whether and how the spread of smartphones enables access and effective use of the internet to reduce poverty [34]. The availability of digital services alone will not have a significant impact unless rural communities' competencies and capacities are improved to use and adopt [28]. Thus Namibia needs to capitalize on the high mobile penetration and improve rural communities' competencies addressing the barriers to use and adopt ICT enabled services to stimulate entrepreneurship activities for employment creation and poverty reduction [29]. Unfortunately, the Government of Namibia has not undertaken policy directives and initiatives to encourage rural entrepreneurs to go digital to transform the economy digitally and to achieve immense diversity by making use of ICT enabled services.

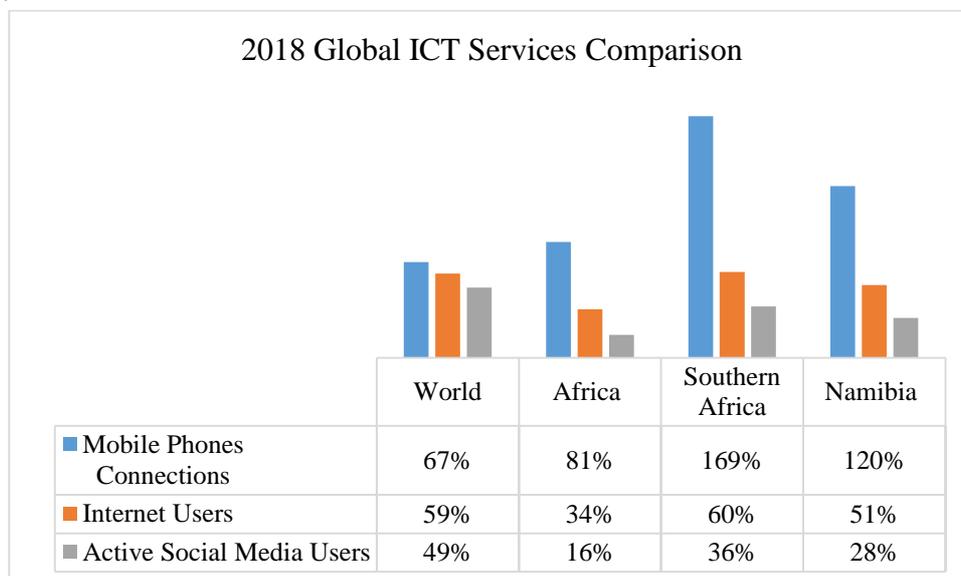

**Figure 1: 2018 Global ICT Services Comparison [11]**

## 2. LITERATURE REVIEW

### 2.1 1CT Enabled Entrepreneurship

There is a consensus that ICT is imperative in national development and that no country can be a player in the global space without digital platforms. However, it is still to be determined whether or more importantly how ICT can work for the marginalised. While ICT is an inevitable tool for an organisation's operations, the marginalised can leverage ICTs in particular through the usage of mobile phones [25]. For example, Tanzanian rural smallholder farmers have used cell phones to obtain agricultural information, transfer money and conduct communications [24]. Harris [5] contends that ICT for development and developmental technology projects failed the marginalised by being irresponsive to their needs and the challenges they face. ICT is perceived as an isolated sector and not yet mainstreamed in developmental agendas and has the minutest funding. However, ICT is crosscutting all the sectors as an enabler for development. Walsham [34] postulates that the new ICT-enabled models can transform the dynamics of the developmental trajectory through digitalisation. However, ICT4D has also been widely criticised as to who derives the benefits, who determines the agenda, and who defines what development means [25]. The main criticism is that the approaches for ICT4D are top-down with limited involvement of the communities and little consideration for their needs and aspirations [35]. It can be opined that we can't decouple ourselves from the digital age but there is a need to contextualise ICT strategies for development to maximise the benefits offered by ICT in particular to rural entrepreneurs. Parida et al. [24] are optimistic that digitalisation eliminates traditional challenges brought about by geographical or physical factors that disadvantage rural entrepreneurs. ICT enabled services can compensate for the remoteness and isolation of rural entrepreneurs and mainstream them into the digital economy. According to Savira and Fahmi [28], rural entrepreneurship is a catalyst for rural development as it creates new opportunities for employment creation, increases community income and diversifies the rural economy. Social and digital entrepreneurship concepts are manifested in developing strategies to enhance internal economic activity through the provision of online banking, e-commerce and adoption of ICTs in private agricultural supply chains which lead to better provision of financial services to citizens and businesses, and improved efficiency in enterprises operations [34]. The emergence of digital entrepreneurship is captured in many studies of [2,6,24 30] which highlights the benefits of entrepreneurs adopting ICT enabled services or using the internet optimality which include: increase sales, cost reduction, high productivity and enhanced quality of goods and services.

## 2.2 Covid-19 effects

COVID-19 had ripple effects with devastating impacts on global economies, supply chains and especially social entrepreneurs' businesses were unfavourable due to border closures and lockdowns [32]. The rural-based SMEs were the most affected. The resilience of entrepreneurs is pivotal to cope and adapt during the COVID-19 devastating period. Resilience is an essential concept in entrepreneurial crisis management as it aids in understanding how businesses adapt or resist change and are, therefore, gaining much attention from academics [32]. For example, 73 entrepreneurs in the UK adapted their business plans successfully and took on new business opportunities during the lockdown by developing new products and services in the digital space [32]. Entrepreneurs who embraced digital services persisted and some even flourished while the ones without digital capabilities perished like the small businesses located in rural areas especially in developing countries. The COVID-19 pandemic has changed the business climate across the world. The significance of digital platforms is immersing, with an increase of online retail sales from 19.4% to 24.6% between August 2019 and August 2020 in China's online market, in Kazakhstan retail sales increased from 5% in 2019 to 9.4% in 2020 and Thailand massive downloads of shopping apps rose 60% in just one week during March 2020 [33]. With the uptake of e-commerce across regions, many consumers and businesses in the global south haven't capitalized on pandemic-induced e-commerce opportunities due to persistent barriers of costly broadband services, lack of digital skills, low income and low interest in e-commerce [37]. The reality of the new normal is underscored by digital entrepreneurship as an alternative to do online business with limited contacts between buyers and sellers. Social entrepreneurship on the other hand played a pivotal role in the provision of food and shelter to marginalised communities during the COVID 19 pandemic [3]. For example, in Namibia, guesthouses and hotels were used as shelters for the homeless and quarantined people and local gardening projects provided for immediate food supplies.

## 2.3 Rural Adoption of ICTs

Adoption of ICTs in rural areas has shown to be more complex considering the conditions of digital infrastructures, but also the socio-cultural and educational context. SMEs capacity to adopt ICT enabled services is undermined by the lack of financial resources and lack of ability to recognize and optimize the benefits that ICT enabled services may offer to the business [9]. A lack of awareness by the underprivileged population about digital services and their potential for sustainable development requires awareness-raising interventions [15]. Small-scale farmers need training as they lack expertise in using cell-phone devices effectively for businesses [17]. Digital literacy was found to be the main challenge in creating opportunities for self-reliance and self-employment [4]. The conceptualisation of digital services that can support new or existing rural businesses has equally shown to be a challenge, in terms of developing business plans encompassing digital marketing for example the absence of 24/7 affordable Internet access, online payment facilities, and English language barriers excluded rural entrepreneurs' full participation on digital marketing space [14]. In summary, studies have revealed that low or lack of ICT adoption can be attributed to a variety of reasons including infrastructure, skills deficiency, cost, limited awareness of benefits, language barriers and an uncertain environment. Some of the barriers can be reduced through policy and regulatory interventions that provide incentives for the operators to invest in rural areas which are less profitable [22].

# 3. METHODOLOGY

## 3.1 Design and Sample

The study adopted a cross-sectional survey of 77 respondents comprising 14 rural entrepreneurs and 63 rural community members from four villages (sites). The selected respondents were regarded as active role players (Community) and key informants (Entrepreneurs). Two questionnaires, one for entrepreneurs and another for non-entrepreneurs were used for the survey that was conducted from August to November 2020 at Okomubonde (Omaheke region), Otumborombonga (Otjozondjupa region), Oniipa (Oshikoto region) and Bordeaux (Hardap region) sites in Namibia. The four sites were selected due to their geographical distribution representing the East, West, North and South of Namibia; moreover, these sites were constructed in 2019 with 3G networks, thus the communities had less than a year of online experience using the internet. A non-probability convenience sampling technique was used to select the participants subject to their availability and willingness to participate in the study.

## 3.2 Study Location

**Bordeaux**
Bordeaux is situated in the Hardap Region and the tower was constructed at a commercial farm. There are very few people living in the area and the majority are farmworkers. The main economic activities are livestock and pig farming. The site is privately owned and there are no entrepreneurship activities found, thus no entrepreneur was interviewed.

**Okomumbonde**
Okomumbonde is located in the Northeast of the Otjinene constituency. The mobile tower also connects Okamuina and Ozongaru communities which are respectively five and seven kilometers away from the site and the survey included all three areas. The main economic activity is agriculture with a focus on small-scale farming with both livestock and crop production. Okomumbonde has no schools, a clinic or a supply of grid electricity. The water supply is from the borehole with a diesel engine water pump. The community has an organized water point committee that collects fees per head of livestock to maintain the borehole, the engine and expenses such as buying of diesel, servicing of the engine and paying the caretaker

**Oniipa**
Oniipa is an old place founded in 1872 by the Finnish Mission and located north of Namibia in the Oshikoto constituency. The town was administered by the Ondonga Traditional Authority until 2004 when it was declared a settlement area and administered by the Oshikoto Regional Council. In 2015, Oniipa was upgraded to a town and constituted its Town Council. There are also several small entrepreneurs operating such as carwashes, salons and food outlets. Besides the MTC tower, there is also a Telecom Namibia mobile tower. The town has a computer training centre which is fully functional and a hospital.

**Otumborombonga**
Otumborombonga is a village situated in Otjozondjupa in the Okakarara constituency. The main economic activity is agriculture with a focus on small-scale farming of both livestock and crop production. The village has about 150 homesteads with a daycare and a gas station.

## 3.3 Data collection instrument and procedures

Data was collected using a researcher-administered questionnaire with statements anchored on a five-point likert scale. The respondents were briefed about the purpose of the study and consent was obtained. They were informed that their participation in the study was voluntary and the information collected would be used for the study and will be treated as confidential. The researcher administered the questionnaire by posing questions/statements to the respondents, listening to the responses and recorded them on the questionnaire which had five possible choices for each statement (1 = strongly disagree, 2 = disagree, 3 = neutral, 4 = agree and 5 = strongly agree). The survey was conducted in English, Otjiherero, or Afrikaans, as chosen by the informant. All questions required to be answered. Besides obtaining demographic data of sex, age and employment status, two guiding questions were posed to the respondents: *What could prevent you from using digital services?* and *What could you propose as an intervention(s) to overcome the problems to adopt digital*

......

*services*. The two questions of "factors preventing adoption" had 10 items while "interventions to adopt digital services" was measured using 7 items. Items were adapted from the previous studies of [ 1,2,9,18]. The responses on items used in the measurement of the variables of the study, "what *prevents* you from *(barriers to)* using digital services" displayed in Table3A, and "*interventions* to adopt digital services" displayed in figure 2 were coded on a five-point Likert scale, ranging from strongly disagree (1) to strongly agree (5).

**3.4 Data Analysis**

In the data interpretation, the rating for proposed interventions was trimmed and responses rating 1-2 were regarded as disagreement while responses rating 3 are neutral responses and rating from 4-5 were regarded as agreement. A descriptive statistic for all ten variables was examined using the Statistical Package for Social Sciences software (SPSS) that was used to create inferences to the research participants. Descriptive statistics are statistics that can be used to describe variables or generalize information from a sample and SPSS software is a widely accepted package for analysing in social sciences [16]. Therefore, to reduce the likelihood of presenting misleading results it is a good practice to use descriptive statistics using a systematic approach [12]. The main purpose of this analysis is to know to what extent is the adoption of ICT enabled services influenced by the ten independent variables and that those measures should be taken based on the results obtained using SPSS. In addition, a multiple regression analysis was done to measure the relationships between independent and dependent variables and to test for respondents' different answers to the research questions. This analysis is suitable to presume the significance of one dependent against two or more independent variables and to provide a prediction about the dependent variable based on its covariance with all the concerned independent variables [16].

## 4. FINDINGS

**Distribution of respondents**

The distribution of respondents according to the region, gender (male/female) and category (entrepreneur/ non-entrepreneur) are presented in Table 1.

| Sites | Entrepreneurs | | Non Entrepreneurs | | Total |
|---|---|---|---|---|---|
| | Female | Male | Female | Male | |
| Bordeaux | 0 | 0 | 1 | 6 | 7 |
| Okomumbonde | 1 | 4 | 9 | 11 | 25 |
| Oniipa | 4 | 2 | 15 | 10 | 31 |
| Otumborombonga | 2 | 1 | 6 | 5 | 14 |
| **Total** | **7** | **7** | **31** | **32** | **77** |

**Table 1: Number of participants**

The respondents' characteristics consist of sex, age group, and employment status of the respondents. As seen in Table 1, the majority of respondents were male representing 51% while female respondents were 49%. Most respondents belonged to the age group of 26-40 (53%) while respondents in the age group of 18-25 were 18% while those in the above 40-years age were 18%. It was found that 53% of the respondents were employed while 47% were unemployed. The majority of respondents are from the age group of 26-40 years and this is due to their active participation in economic activities and exposure to modern technologies. This finding is consistent with the previous study outcome that due to the novelty of technology the younger individuals are presumed to be more inclined to access the internet than older ones [15].

**Factors influencing ICT Adoption among Rural Entrepreneurs in Namibia**

Descriptive Statistics and Multiple linear regression were used to assess the factors that influence the adoption of ICT enabled services and the result is presented in Table 2A and 2B respectively.

| Descriptive statistics | | | | | | | | |
|---|---|---|---|---|---|---|---|---|
| **Variable** | **N** | **Mean** | **Median** | **Mode** | **SD** | **Range** | **Min** | **Max** |
| Lack of Awareness | 77 | 4.26 | 5 | 5 | 1.03 | 4 | 1 | 5 |
| Social Influence | 77 | 1.97 | 1 | 1 | 1.38 | 4 | 1 | 5 |
| High cost to connect to the internet | 77 | 3.94 | 4 | 4 | 1.02 | 4 | 1 | 5 |
| Lack of Access to Smart devices | 77 | 4.05 | 4 | 5 | 1.18 | 4 | 1 | 5 |
| Lack skills to use the internet | 77 | 3.90 | 4 | 4 | 0.88 | 4 | 1 | 5 |
| Lack of electricity supply to charge the devices | 77 | 4.14 | 4 | 5 | 0.97 | 4 | 1 | 5 |
| Security concerns / Safety of internet | 77 | 3.53 | 4 | 4 | 0.97 | 4 | 1 | 5 |
| Limited access to high-speed internet | 77 | 3.58 | 4 | 4 | 0.99 | 4 | 1 | 5 |
| Limitation to purchase recharge services | 77 | 3.47 | 3 | 3 | 0.91 | 4 | 1 | 5 |
| Limited useful local content | 77 | 3.29 | 3 | 3 | 0.86 | 4 | 1 | 5 |

**Table 2A: Factors influencing ICT Adoption among Rural Entrepreneurs in Namibia**

Looking at descriptive statistics for the 10 variables above, respondents mostly agreed with "Lack of awareness" to be their major influence preventing them from using digital services/internet. Respondents mostly disagree with "Social Influence" to be the major influence that could prevent someone from accessing digital service/Internet. With a combined percentage disagreement of 70.1% and total participants of 54 out of 77.

| Coefficients | | | | | | |
|---|---|---|---|---|---|---|
| Model | | Unstandardized Coefficients | | Standardized Coefficients | t | Sig. |
| | | B | Std. Error | Beta | | |
| 1 | (Constant) | 0.874 | 0.373 | | 2.342 | 0.022 |
| | High cost to connect to internet | 0.068 | 0.064 | 0.102 | 1.072 | 0.288 |
| | Lack of Access to Smart devices | 0.097 | 0.053 | 0.172 | 1.822 | 0.073 |
| | Lack skills to use internet | 0.262 | 0.07 | 0.34 | 3.717 | 0 |
| | Lack of electricity supply to charge the devices | 0.041 | 0.075 | 0.058 | 0.545 | 0.588 |
| | Security concerns / Safety of internet | -0.086 | 0.063 | -0.122 | -1.366 | 0.176 |
| | Limited access to high speed internet | 0.011 | 0.072 | 0.015 | 0.145 | 0.885 |
| | Limitation to purchase recharge services | 0.055 | 0.08 | 0.073 | 0.68 | 0.499 |
| | Lack of awareness | 0.309 | 0.059 | 0.468 | 5.199 | 0 |
| | Social Influence | 0.046 | 0.041 | 0.093 | 1.125 | 0.265 |
| | Limited useful local content | 0.021 | 0.077 | 0.026 | 0.277 | 0.783 |

**Table 2B: Factors influencing ICT Adoption among Rural Entrepreneurs in Namibia**

The model linking factor limiting the use of digital/internet services to promote adoption was significant with R-square = .604, p <0 Lack of skills and lack of awareness have a very significant influence on adoption (p <0); while access to smart devices significantly influences adoption at p <.1 (p=.073).

### Alternatives to reduce barriers to ICT adoption

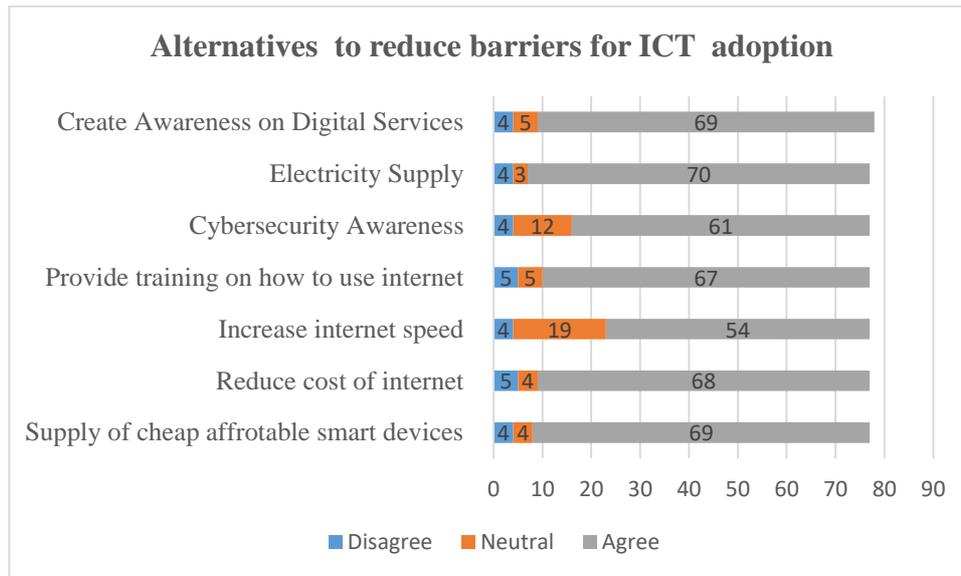

**Figure 2:  Alternatives to reduce barriers for ICT Adoption**

As seen from Figure 2, the majority 70 (90.9 %) of respondents proposed electricity supply, creation of awareness on digital services 69 (89.6%) while 68 (88.3%) proposed reduction of cost of the internet; 67 (87%) proposed the provision of training on how to use internet and 61 (79.2 %) proposed the creation of Cybersecurity awareness.

## 5. DISCUSSION

The study found that the low adoption of ICT enabled services by rural entrepreneurs is influenced by a lack of awareness of digital services, Lack of skills, lack of electricity supply and lack of access to smart devices and Cybercrime.  Hence the adoption of digital services is not optimally realised to stimulate entrepreneurship activities for socio-economic development. Conceptually, ICT enabled services adoption drives generativity, creates value chains of products and services and enables borderless participation of geographically dispersed audiences and enables entrepreneurs to connect with customers and producers; cyberspace creates some fair market participation opportunities to all participants in the economy [8]. Notably, besides the various benefits deriving from using digital services, it is presumed that rural communities are not optimally using available digital services to promote and stimulate entrepreneurship activities but mostly, they use them for communication and social media purposes [18]. Conversantly, the decision to either use ICT application or not is influenced by various factors which include economic, technical, political and social [9]. Namibia has a high mobile phones penetration rate and broadband policy initiatives to connect rural areas but it lacks a policy directive that promotes rural digital entrepreneurship.

The study conducted by Sanga and Buzingo [27] contends that for people to make use of the technologies, they have to be aware of them. In this regard, there is no ambiguity that awareness is considered as one of the major constraints of ICT adoption among rural-based SME entrepreneurs [17]. Musingafi and Zebron [22] argue that the selection of the type and use of ICTs must be based on appropriateness to the needs and expectations of the end-user. Due to a lack of awareness of online marketing, online shopping is low and this will be detrimental to rural entrepreneurship activities as the COVID-19 pandemic dictates limited face-to-face interventions. Therefore, if rural enterprises continue not using digital platforms, they will face challenges to reach the target markets and maximise the sales of goods and services. It has been argued that the active participation of communities is required for the successful adoption and use of ICTs, however not just as beneficiaries but as people who need to trust and use services and technology tools. Kyobe [17] highlighted that the lack of power supply has remained a major problem in Africa and predominantly in rural areas. Thus, ICT devices are reliant on electricity, and the shortage of electricity supply causes disruptions in operations [4]. The power supply is a key facet of digitalisation transformation. We also found that lack of know-how to use smart devices and to browse on the internet limit the ability of users to connect to the internet and this also contributes to cybersecurity fears as a rural community is not able to apply the necessary cybersecurity measures. This concurs with the findings by [2,21] that ICT skill deficiencies among rural communities impede the usage hence low adoption as users will adopt the technology if it is easy to use because these skills are a prerequisite for usage. The study by [1] found that there was a strong positive correlation between ICT knowledge and skills and ICT adoption. Most rural communities are unemployed or have low income thus their ability to procure smart devices and data is limited. We contend that a high adoption rate of digital entrepreneurship in rural areas will nurture employment opportunities and increase the income of rural communities. It will also serve as motivation for young entrepreneurs to stay in rural areas and actively participate in rural development.

### 5.1 Conceptual ICT Adoption Model

From the empirical data collected, five key barriers to ICT enabled applications and services were identified and possible interventions are proposed. In light of the factors influencing the adoption of ICT by rural entrepreneurs, a conceptual framework has been developed and is shown in Figure 3. This suggests some possible interventions that could promote ICT enabled services adoption in rural areas. To the end, actors are identified to ensure that the proposed interventions are implemented to reduce the barriers to ICT adoption. The roles and responsibilities of each actor are outlined in the recommendations.

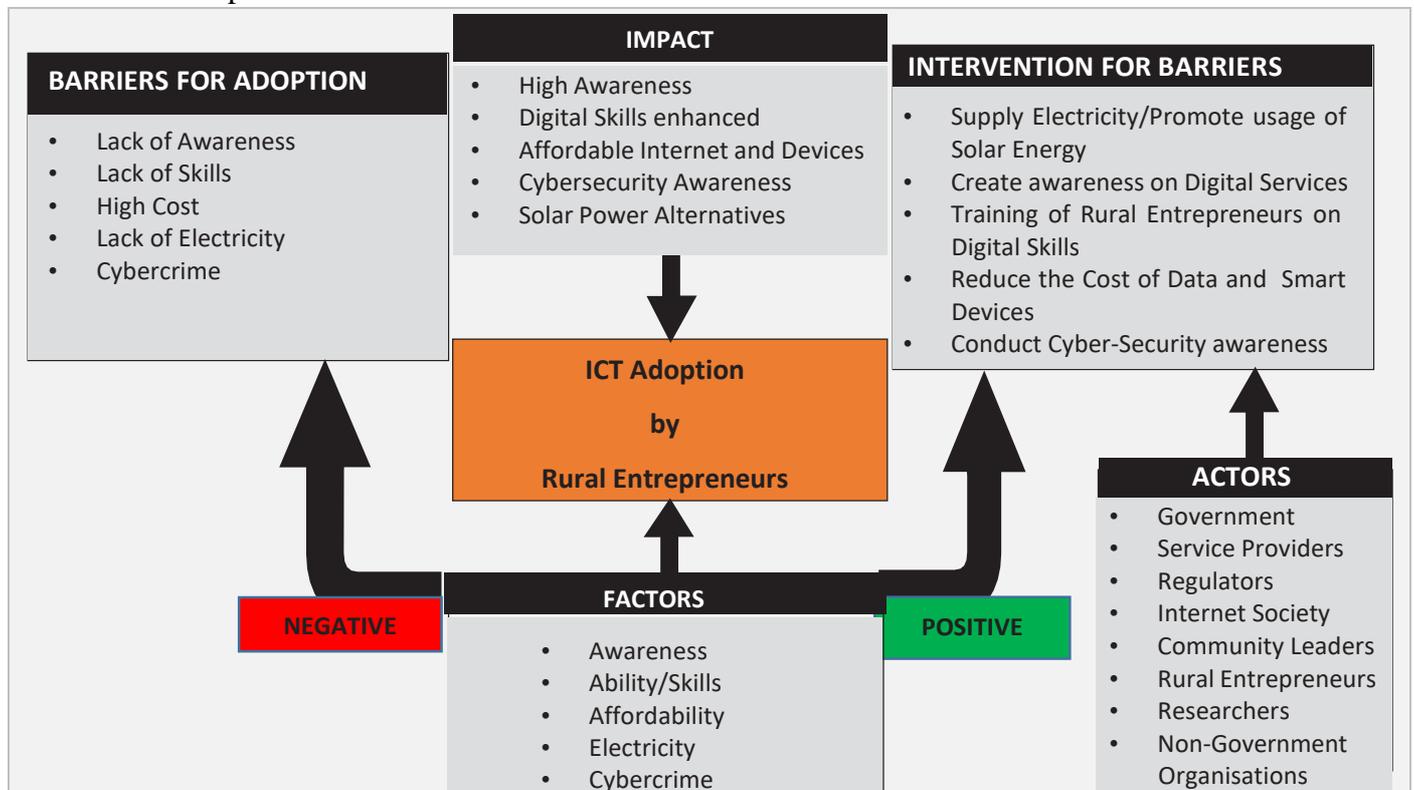

**Figure 3: Proposed Conceptual ICT adoption model for rural entrepreneurs**

## 6. CONCLUSION AND RECOMMENDATIONS

### 6.1 Conclusion

Rural entrepreneurs play a pivotal role in Namibian's economy in employment creation and mitigation of the rural-urban migration. This study found that rural entrepreneurs in Namibia are facing the same challenges as per the findings of other studies of [4,21,38]. For this study, the most dominant factors are a lack of awareness of the importance of ICT, lack of skills, lack of electricity and cybercrime. A conceptual model for these five main factors is proposed with alternative interventions for possible mitigation of the identified challenges. Conversely, the model will only be feasible if interventions are undertaken to address the observed challenges. These factors are similar to findings in other developing countries as stated by Kyobe [17] that in developing nations the adoption of ICT is mainly influenced by the income and wealth of an individual, its computer and internet skills and infrastructure. The digital entrepreneurs have prospects and are a solution in embracing the future and using ICTs for boosting business growth will be pivotal for rural entrepreneurs to recover from the impacts of COVID-19. In this paper, we have argued that ICT provides information about access to market and business, brings financial services literally to the fingertips of rural consumers, enhances social cohesion and creates collective and shared platforms for the communities to exchange know-how and ideas. We thus conclude that high-quality internet provision can help unlock the potential of rural areas, and can make them more attractive places to live. During COVID-19, ICT systematically enabled the connection of people to people, business to customers and government to citizens.

As stated by Lwago [20] that many studies assessed the contribution of ICT for development in diverse perspectives such as poverty reduction, sustainable livelihoods, economic growth, and the view of development as freedom. The new knowledge can lead to improved utilization of ICT enabled services and addressing socio-technical challenges for adoption. This study mapped the contribution of ICT to development, by assessing the factors influencing the adoption of ICT enabled services while providing insights to the developmental agencies, government and service providers to take cognizance of pertinent factors that could influence the optimal usage of available ICT enabled services by the community and reaping the associated benefits. The study has some significant implications for rural-based entrepreneurs to make use of benefits derived from ICT enabled services such as improved marketing strategy. This study is a value-addition to the existing information systems literature in three ways of providing a succinct account of the impact of digital platforms in rural settings; provides a future research agenda for digital entrepreneurship in the context of developing countries and provides a deep understanding of the ICT4D literature in respect of linkage between development, technology and people.

Limitations include a rural communities sample focusing on entrepreneurs. The survey was only conducted at four sites, so findings have limited generalisation but can be used to build some hypotheses for the other areas. Future research should devote more attention to more entrepreneurs in both rural and urban areas with a specific target on youth and women entrepreneurs. Furthermore, a bigger size sample size is recommended for more reliable and valid results. There is a need to implement the possible interventions to test and validate the proposed conceptual ICT adoption model for rural entrepreneurs.

## 6.2 Recommendations

Based on the finding that there is low usage of ICT enabled services for entrepreneurship activities by rural communities especially online marketing and online shopping, there is a need for the Namibian government and Internet Society to promote the use of the internet to benefit entrepreneurship activities in rural areas. The stakeholders need to and design fit for purpose digital services and develop need-specific applications for rural entrepreneurs that are simple and easy to use. Winschiers-Theophilus et al, [35] advocate for a consultative and participatory approach before and during the deployment of projects to rural areas. This enables the affected communities to provide their needs and aspirations and the bottom-up approach will inter alia increase awareness before the deployment of technology and enhance the motivation for the adoption of such technology. For a meaningful development of suitable ICT enabled services, the real needs and requirements are to raise awareness about ICT to the population that lives under difficult conditions in remote locations [21]. Considering the low or lack of skills by rural communities to use smart devices and to navigate the internet for socio-economic purposes. There is a need for the government, service providers and Internet Society and other stakeholders to initiate programmes that will lead to enhanced capacity of the rural community digital skills and literacy. The training initiatives should also include raising awareness of available digital services and cybersecurity awareness, the latter will enlighten the rural entrepreneurs of the associated benefits of using the internet for business purposes and equally equip them with essential cybersecurity measures to protect them when online. We suggest proper coordination between service providers and the Ministry of Mines and Energy to prioritise the rollout of rural electrification to the sites that are planned for the construction of the mobile tower. It is recommended that electricity should be made available to the sites before mobile networks construction to have a high power mobile network for better internet speed and constant electricity supply to the community to charge the devices. Lastly, the Government should also consider reducing the import taxes on ICT devices that will eventually reduce the price of devices to the end-users.